\documentclass[11pt]{article} 
\usepackage{hyperref} \pdfoutput=1
\usepackage{graphicx}
\usepackage{upgreek}

\begin{document} 
\title{Numerical simulation of blood with fluid-structure interactions using the lattice-Boltzmann method}
\author{Daniel A. Reasor Jr., Jonathan R. Clausen, Brian M. Yun, and Cyrus K. Aidun\\ \\\vspace{6pt} George W. Woodruff School of Mechanical Engineering, \\ Georgia Institute of Technology, Atlanta, GA 30332, USA}

\maketitle

\begin{abstract} 
	The fluid dynamics video presented here outlines recent advances in the simulation of multiphase cellular blood flow through the direct numerical simulations of deformable red blood cells (RBCs) demonstrated through several numerical experiments. Videos show RBC deformations in variety of numerical simulations, relative viscosity of a suspension of RBCs in shear, and the cell-depleted wall layer for blood Hagen--Poiseuille flow.
\end{abstract}

The animations presented in the submission are results from the direct numerical simulation of deformable RBC suspensions using a hybrid lattice-Boltzmann spectrin-link (LB-SL) method\cite{Piv:08} that is coupled with a particle--particle and particle--boundary contact model of MacMeccan et al.\cite{Mac:09}. The implementation is based on the parallel framework of Clausen et al.\cite{Cla:10} has also been optimized to run on as many as 65,536 cores of the IBM \emph{Blue Gene/P} architecture. In order to simulate high volume fraction (hematocrit) simulations of deformable particles in wall-bounded or cylindrical domains, a seeding method is used. In this procedure, the particles are initialized at 30\% of their final size and grown to their full size rigidly, allowing particle--particle and particle--wall interactions using the lubrication and contact model previously referenced. Each RBC is composed of a triangulated spectrin network consisting of $N$=613 nodes. Physically, RBCs are biconcave in shape with a maximum diameter of $8\,\upmu$m enclosed by an elastic membrane with an effective elastic shear modulus of $5.7\times10^{-3}\,\mbox{dyn cm}^{-1}$~\cite{Wau:79} and a bending stiffness of $2.2\times10^{-12}\,\mbox{dyn cm}$~\cite{Hwa:97}, which surrounds a liquid hemoglobin with a viscosity of 6~cP. The RBCs are suspended in blood plasma with a viscosity of 1.2~cP. The relative non-dimensional elastic parameter is the elastic capillary number, $\mbox{Ca}_G \equiv \mu \dot{\gamma} a/G_m$ where $\mu$ is the fluid viscosity, $\dot{\gamma}$ is the local shear rate, $a$ a length scale based on the particle size, and $G_m$ is the elastic shear modulus. 

Isolated RBC simulations are first presented. These simulations include a numerical experiment chosen to mimic an optical tweezer experiment. For isolated RBCs in shear, two different dynamic regimes are then shown demonstrating the tumbling versus tank-treading phenomena. A parachuting RBC is then animated in a microvessel sized rigid tube. The flow in that simulation is driven by an axial body force used to imitate a constant pressure-gradient. 

Dense suspensions of RBCs follow starting with a demonstration of the non-Newtonian effect known as shear thinning is then given by results of simulations using the Lees-Edwards boundary condition (LEbc) for a suspension of spectrin-link based RBCs. As the hematocrit (or volume fraction) of RBCs increases, so does the relative viscosity. This can be shown using the same simulations used to demonstrate shear thinning with the LEbc. The cell depleted wall layer is then highlighted through an animation of a microvessel sized tube with a focused frame near the tube wall. Ensemble averaged hematocrit and velocity are also given to demonstrate the cell depleted layer thickness and the blunting of the velocity profile due to the presence of RBCs. The accumulation of shear stress on the surface of platelets is one way to model ``blood damage'' through blood damage index (BDI) models. By simulating a dense suspension of RBCs with a realistic number of platelets we can monitor the effects of RBC-platelet interactions and the effect on the platelet shear stress. 

\subsection*{Acknowledgements}
D.R. is funded by the U.S. Department of Defense through the SMART fellowship program and J.R. was funded by the Institute of Paper Science and Technology at Georgia Tech. These simulations were performed through the use of TeraGrid Queen Bee Linux Cluster at LONI at Louisiana State University made available by the National Science Foundation. 

\bibliographystyle{unsrt}		% BIB Style

\end{document}